# WS$_2$ Monolayers Coupled to Hyperbolic Metamaterial Nanoantennas: Broad Implications for Light−Matter-Interaction Applications

S.R.K.Chaitanya Indukuri[†], Christian Frydendahl, Jonathan Bar-David, Noa Mazurski and Uriel Levy[†]

Department of Applied Physics, the Faculty of Science and the Center for Nanoscience and Nanotechnology, The Hebrew University of Jerusalem, Jerusalem, 91904, Israel.

[†] Corresponding Author: stiaram.indukuri@mail.huji.ac.il, ulevy@mail.huji.ac.il

**Abstract:**

Due to their atomic layer thickness, direct bandgap, mechanical robustness and other superior properties, transition metal dichalcogenides (TMDCs) monolayers are considered as an attractive alternative to graphene for diverse optoelectronic applications. Yet, due to the very nature of atomic layer thickness, the interaction of light with TMDCs is limited, hindering overall efficiency for optical applications. Therefore, in order for TMDCs to become a true candidate as the material of choice for optoelectronics, there is a need for a mechanism which significantly enhances the interaction of light with TMDCs. In this paper, we demonstrate about 30-fold enhancement of the overall photoluminescence emission intensity from a $WS_2$ monolayer, by its coupling to a hyperbolic metamaterial nanoantenna array. This enhancement corresponds to nearly 300-fold enhancement per individual nanoantenna. This overall enhancement is achieved by the combination of enhancing the excitation (absorption) efficiency, alongside with enhancing the radiative decay rate. Our result paves the way for the use of TMDCs in diverse optoelectronic applications, ranging from light sources and photodetectors to saturated absorbers and nonlinear media.

**INTRODUCTION:**

Transition metal dichalcogenides (TMDCs) are direct bandgap semiconductors at the monolayer limit[1], exhibiting bright excitonic photoluminescence at visible frequencies with a great promise for next-generation photonic applications[2]. Unfortunately, low intrinsic quantum efficiency of emission and weak absorption limits their applicability in current state of the art photonic systems and applications. Yet, their atomic scale thickness and compatibility with different substrates, makes these materials highly attractive for integration with nanoscale photonic components with the goal of significantly enhancing their emission rate[3,4]. And indeed, the integration of monolayer TMDCs into different photonic components like photonic crystals and nanobeam waveguides has already been demonstrated[5–7]. In such cavities, high quality factors (Q-factors) can be achieved alongside with diffraction limited mode volumes at the wavelength scale.

While dielectric structures show promising results, further reduction in cavity sizes and mode volumes of dielectric cavities is limited by diffraction, which is governed by the dielectric constant of the cavity materials. This restriction can be removed by using noble metal nanostructures supporting localized surface plasmon polaritons (SPPs). In such a case, the effective mode volume can be further reduced down to deeply sub-wavelength sizes[8–11]. The spontaneous emission rate of emitters coupled to such structures is increased by the Purcell factor $F_P$, which is proportional to $\frac{Q}{V_m}$ where $Q$ is the quality factor and $V_m$ is the mode volume of the cavity. In the case of dielectric photonic structures $Q$ can become very large whereas the smallest $V_m$ is on the order of $\left(\frac{\lambda}{n}\right)^3$ (with $n$ being the materials refractive index), and thus the corresponding $F_P$ values are moderate. Furthermore, placing the emitter in close proximity to the cavity may hamper the achievable Q factor values. In contrast, plasmonic structures support further reduction in mode volume down to the deep subwavelength level, with potential increase in $F_P$ and large enhancement of emission rate[12]. Unfortunately, this enhancement due to the Purcell factor is limited by the nonradiative contribution (also known as quenching) originated from the Ohmic loss in the metal. Insight into the physical nature of this limited enhancement indicates that the Purcell factor is spectrally similar to the scattering cross-section of the plasmonic particle, whereas the nonradiative decay rate of the plasmonic particle is spectrally similar to the absorption of the plasmonic particle[13,14]. Unfortunately, for most of the plasmonic structures used for enhanced emission of TMDCs both the scattering- and absorption cross-sections overlap spectrally, due to the intrinsic nature of the plasmonic resonance itself. In such a case, the total decay rate enhancement (Purcell factor) is mostly dominated by the enhancement of the nonradiative decay rate, and thus the overall Purcell enhancement in plasmonic structures is limited.

In order to significantly enhance the emission collected from an atomically thin layer emitter which is coupled to a plasmonic structure, one needs to increase both the excitation efficiency (i.e. the absorption of the excitation light which is typically very small due to the extremely low thickness of the layer) and the internal quantum efficiency. The latter is governed by the radiative decay rate and the nonradiative decay rate. Unfortunately, enhancing the radiative decay rate comes at the expense of enhancing the non-radiative rate due to the spectral overlapping of absorption and scattering cross section. Other avenues for enhancing the collected emission is by improving the light collection efficiency (e.g. by controlling the emission angular distribution), and increasing the density of the plasmonic structures per unit area.

To date, a large body of research related to enhancing the emission intensity from TMDCs is focused on using different plasmonic[15–21] and photonic structures[22–26] to achieve large excitation enhancement through hotspot engineering, rather than enhancing the internal quantum efficiency of the TMDCs coupled to the nanostructures.

In this work, we use WS$_2$ as an active medium coupled to hyperbolic metamaterial (HMM) nanoantenna, and demonstrate large enhancement of the overall photoluminescence (PL), up to 30-fold compared to the bare monolayer over the entire area of coverage, and up to ~300 fold per nanoantenna. This is achieved by addressing most of the aspects that have been mentioned before. First, as in previous works, we improve the excitation efficiency. Next, we design and fabricate our HMM nanocavity array in such a way that there is a spectral separation between the radiative and the non-radiative decay channels. This enables enhancement of the internal quantum efficiency of the WS$_2$ emitter. Furthermore, we use a high density nanoantenna array which allows us to obtain large enhancement of the overall PL rather than enhancing the PL from a single nanoantenna (see e.g. the slot antenna coupled to TMDC[27]).

**RESULTS AND DISCUSSION**

Figure 1(a) shows a schematic diagram of the experimental system. Here, a WS$_2$ monolayer is coupled to a hyperbolic metamaterial (HMM) nanoantenna array. HMMs can exhibit unusual optical properties, such as negative refraction[28], perfect absorption[29], enhanced light-matter interactions[30–35], and long-range dipole-dipole interactions[36]. It has also been shown that cavities made of HMM media show unusual scaling laws for electromagnetic field confinement and strong light-matter interactions compared to other regular dielectric cavities[37–39]. Very recently, it was experimentally demonstrated that hyperbolic nanoantenna can be used for pure and spectrally separated highly radiative (scattering) and highly nonradiative (absorption) channels in the visible region[40]. Figure 1(b) shows a schematic diagram of the hyperbolic iso-frequency surface (metamaterials) and spherical iso-frequency surface (dielectrics).

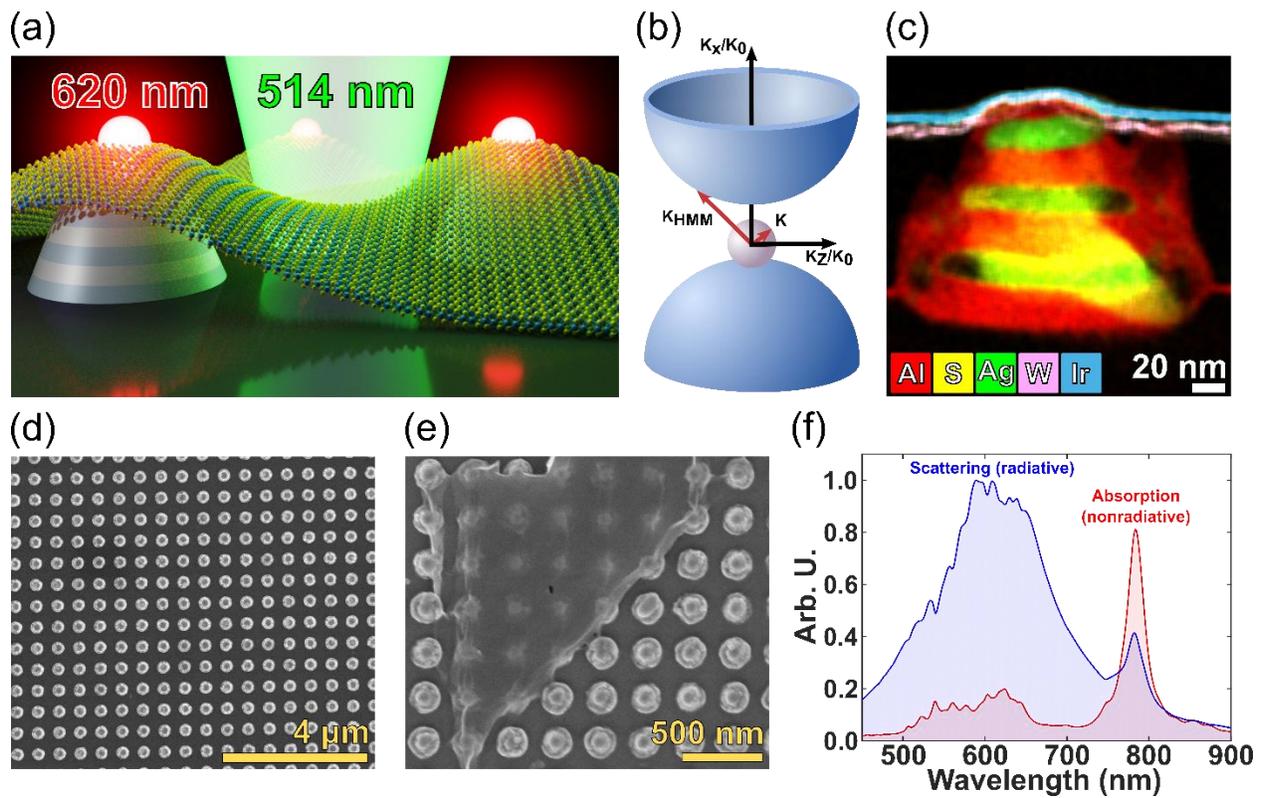

**Figure 1: (a)** Schematic diagram of the experimental configuration. Here the $WS_2$ monolayer is coupled to HMM nanoantenna by placing it on top of the structure. **(b)** Shows the schematic diagram of the hyperbolic iso- frequency surface of the HMM. For comparison, the standard spherical surface of dielectric materials is also shown. (c) STEM image showing a cross section of a $WS_2$ monolayer coupled to HMM nano antenna with chemical composition analysis (indicating the different layers of the HMM and the $WS_2$ monolayer). **(d, e)** SEM top view images of HMM nanocavities (Scale bar 4 μm and 500 nm respectively) fabricated using e-beam lithography, with WS2 flake on top of cavities as shown in (e). **(f)** Normalized calculated scattering and absorption cross-section of the HMM nanocavities with a diameter at the bottom of 160 nm and pitch of 380 nm. Here, the structure is on a glass substrate ($n_{glass}$ = 1.5), and is illuminated with a plane wave from the air side.

We designed a hyperbolic nanoantenna array as follows. Each hyperbolic nanoantenna contains a three period stack of alternating metal-dielectric layers with a circular (slightly conical) shape having a 160 nm diameter. The stack is composed of 16 nm Ag and 24 nm $Al_2O_3$, having a total thickness of 120 nm. All the antenna array is covered with a 3-4 nm atomic layer deposited (ALD) capping layer of $Al_2O_3$ to avoid oxidization of the silver. The nanoantennas are designed to have a hyperbolic dispersion at the emission spectrum of the $WS_2$ monolayer [see SI]. Figure 1(c) is a scanning transmission electron microscope (STEM) image showing a cross section of a $WS_2$ monolayer coupled to HMM nanoantenna with chemical composition analysis (indicating the different layers of the HMM and the $WS_2$ monolayer). Figure 1(d) and (e) show the scanning electron microscope (SEM) images of the HMM nanoantenna array, with a triangular $WS_2$ flake visible in (e). One of the most important optical properties of our HMM nanoantenna is the ability to spectrally separate the pure radiative and non-radiative channels. This is demonstrated in Figure 1(f) showing the normalized scattering and absorption cross-sections for our nanoantenna design. Indeed, from figure 1(f) it is clear that our nanoantenna have two separate spectral regions of interest. The HMM nanoantenna's scattering (radiative channels) is dominant in the spectral region around 600 nm, whereas the spectral band around 800 nm is dominated by the absorption (nonradiative channels). To achieve this spectral separation, the iso-frequency surface needs to be hyperbolic. Without hyperbolic dispersion, these antennas do not show the spectral separation of radiative and non-radiative channels[40]. While the spectral separation of radiative and non-radiative channels in metamaterial nanoantenna has been demonstrated previously, the effect on light-matter interaction due to spectral separation has not yet been shown.

Figure 2(a) shows measured and finite difference time domain (FDTD) calculated transmission spectra of the hyperbolic nanoantenna array. The transmission spectra show two dips at a wavelength around 600 nm and at 820 nm, respectively. These dips correspond to the (2,2,1) and (2,2,2) modes, respectively. They depend upon the number of nodes in the electric field profile. The details of the physical nature of these modes are given in a previous report[39]. The mode (2,2,1) near 600 nm is dominated by pure radiative channels (scattering) whereas the (2,2,2) mode is dominated by the nonradiative channels (absorption), as indicated in the scattering and absorption of these cavities in figure 1(f). The FDTD simulations are in good agreement with the experimental results. Figure 2(b) shows a $WS_2$ monolayer transferred on to the hyperbolic nanoantenna array (light red color). The insert in figure 2(b) shows the same monolayer on $Si/SiO_2$ (285nm) before transfer, and the area of the monolayer is around 10 µm × 20 µm. The monolayer is transferred using a conventional dry transfer process, details are given in the methods section below[41–43].

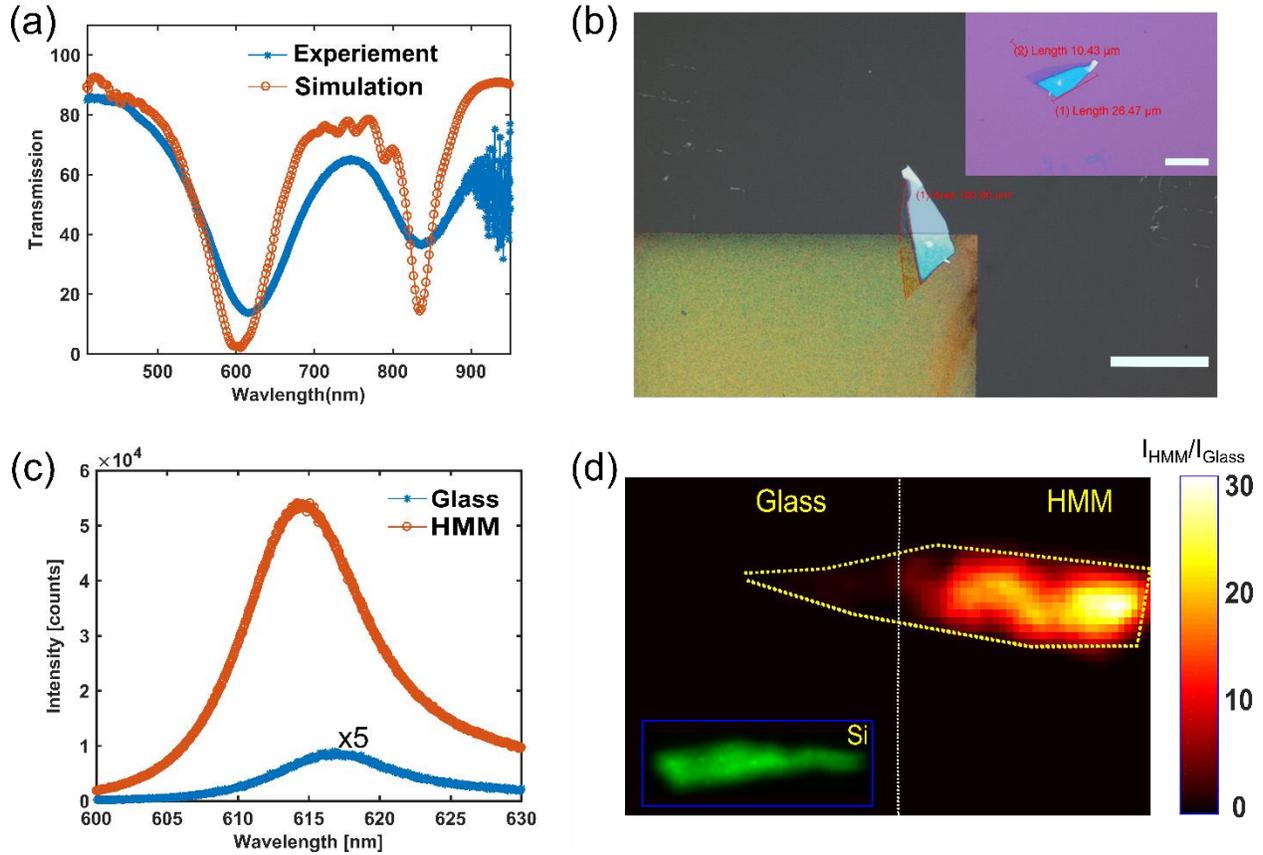

**Figure 2:** **(a)** Experimentally measured (blue), and calculated (by FDTD, red) transmission spectrum of light propagating through the HMM nanoantenna array. **(b)** Optical image of the $WS_2$ monolayer on top of the HMM nanoantenna array. The array has a light greenish color when compared to background glass (black-grey) color. Insert in figure (b) shows the optical image of the same layer on Si/SiO$_2$ (285 nm). **(c)** Enhanced steady-state photoluminescence (PL) from the $WS_2$ monolayer coupled to the HMM nanoantenna array (orange), as compared to the PL emission from the $WS_2$ placed on the glass substrate (blue). **(d)** measured PL emission from the $WS_2$ monolayer coupled to HMM nanoantenna array integrated over the spectral regime of 600-630 nm, normalized to $WS_2$ monolayer on top of the glass. The inset shows the comparative spectral map of the same $WS_2$ monolayer on the Si/SiO$_2$ substrate.

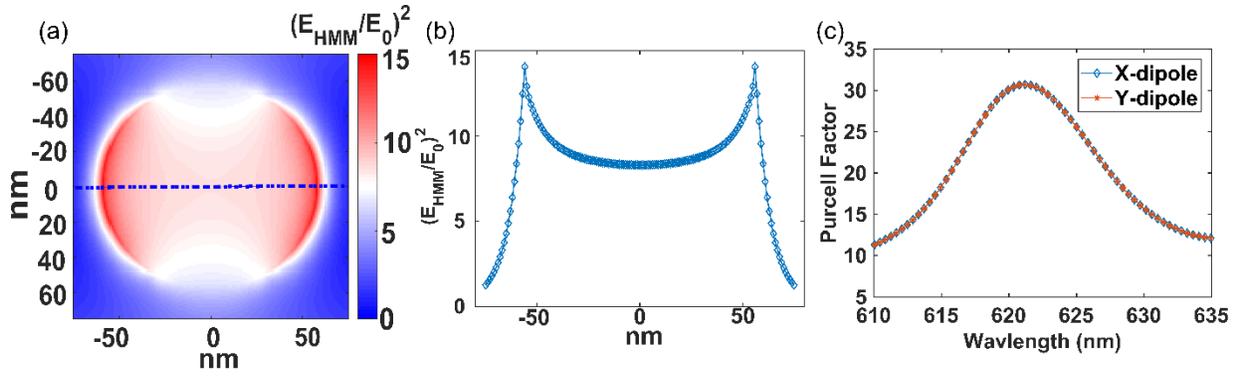

**Figure 3:** FDTD calculations of the emission enhancement from a single nano antenna. **(a)** Normalized electric field intensity enhancement at the excitation wavelength (514 nm) for the electric field distribution at 0.5 nm above the top of the HMM nanoantenna. **(b)** Cross section of the Electric field intensity enhancement distribution along the dashed line of panel (a). **(c)** The Purcell factor at the emission wavelength (620 nm) as a function of wavelength for both x- and y-axis polarized dipoles, calculated at the position of one of the edges of the HMM nanoantenna.

Figure 2(c) shows how the HMM nanoantenna array enhances the PL emission significantly as compared to the PL emission measured from the same layer at a neighboring area on glass. Due to the conical shape of the antenna, its diameter at the top is ~100 nm. As a result, the array has a ~10% effective filling fraction by area. The PL emission from the array is found to be enhanced by a factor of ~30 relative to the emission from the layer on glass. We deliberately transferred the monolayer such that the same monolayer is partially placed on the glass and partially on the HMM cavities. This is the best way to self-consistently compare the PL emission enhancement, as comparing different flakes may not be accurate. Large enhancement is observed over the whole HMM nanoantenna array. We can also calculate the enhancement of PL per nanoantenna by dividing the overall measured enhancement by the filling fraction, resulting in an enhancement per nanoantenna of ~300.

To understand the origin of this enhancement, we performed rigorous FDTD simulations. The total enhancement is the result of three independent processes: 1 – Excitation enhancement (external quantum efficiency), $\frac{\gamma_{\text{HMM}}^{\text{exc}}}{\gamma_0^{\text{exc}}}$, which is calculated at the excitation wavelength (around 514 nm). 2 – emission enhancement, $\frac{\gamma_{\text{HMM}}^{\text{r}}}{\gamma_0^{\text{r}}}$, (Purcell factor or internal quantum efficiency) which is calculated at the emission wavelength (around 620 nm). 3 – enhancement of the collection efficiency, $\frac{\eta_{\text{HMM}}}{\eta_0}$. Due to the excitation being well below the saturation limit, these processes are independent. The total enhancement is the product of these three independent quantities, defined as[26,44,45]:

$$F_{\text{FDTD}} = \frac{\gamma_{\text{HMM}}^{\text{exc}}}{\gamma_0^{\text{exc}}} \cdot \frac{\gamma_{\text{HMM}}^{\text{r}}}{\gamma_0^{\text{r}}} \cdot \frac{\eta_{\text{HMM}}}{\eta_0} \ . \qquad (1)$$

In our FDTD simulations, excitation enhancement is the ratio of the excitation rate of a $WS_2$ monolayer on top of the HMM nanoantenna normalized to the excitation rate of a $WS_2$ monolayer on top of a glass substrate at the excitation wavelength (514 nm). In simulations it is assumed that $\frac{\gamma_{\text{HMM}}^{\text{exc}}}{\gamma_0^{\text{exc}}} = \left|\frac{E_{\text{HMM}}}{E_{\text{glass}}}\right|^2$. The spatial distribution of this enhancement is shown in Figure 3(a). Figure 3(b) shows a line scan of Figure 3(a) along the center of the nanoantenna. As can be seen, the simulations predict excitation enhancement in the order of 10 fold.

The second term is the enhancement of the radiative decay rate which is equivalent to the enhancement of the quantum yield $\frac{Q_{\text{HMM}}}{Q_0}$, and is calculated using the Purcell factor[46–48]. In general, the Purcell factor gives the total decay rate, both radiative and nonradiative decay rate enhancements. The quantum yield of spontaneous emission is $Q = \frac{\gamma_{\text{r}}}{\gamma_{\text{r}} + \gamma_{\text{nr}}}$. For very small internal quantum efficiency ($\gamma_{\text{nr}} \gg \gamma_{\text{r}}$), as is typically the case in $WS_2$, the Purcell factor and quantum yield enhancement are equal to the radiative decay rate enhancement. Thus, we can assume that the total radiative decay rate enhancement is equal to the Purcell factor. As shown in figure 3(c), the radiative rate enhancement is around ~30 per single cavity at peak enhancement is achieved at the WS$_2$ emission wavelength of ~620 nm. In this calculation, WS$_2$ is modeled as a dipole whose emission spectrum matches with the PL emission of the WS2 monolayer. For both x- and y-polarization, the Purcell factor is the same almost at any point on top of the HMM nanoantenna. More details of these calculations given in methods and in the supplementary materials.

The final term in equation 2 is about 1 for our case (see Fig. SI4 for more details). This is because our current HMM nano antenna array does not enhances the directionality of the emission from the WS$_2$ monolayer coupled to HMM nanoantenna as compared with the WS$_2$ monolayer on the glass substrate. This can be a point for future improvement.

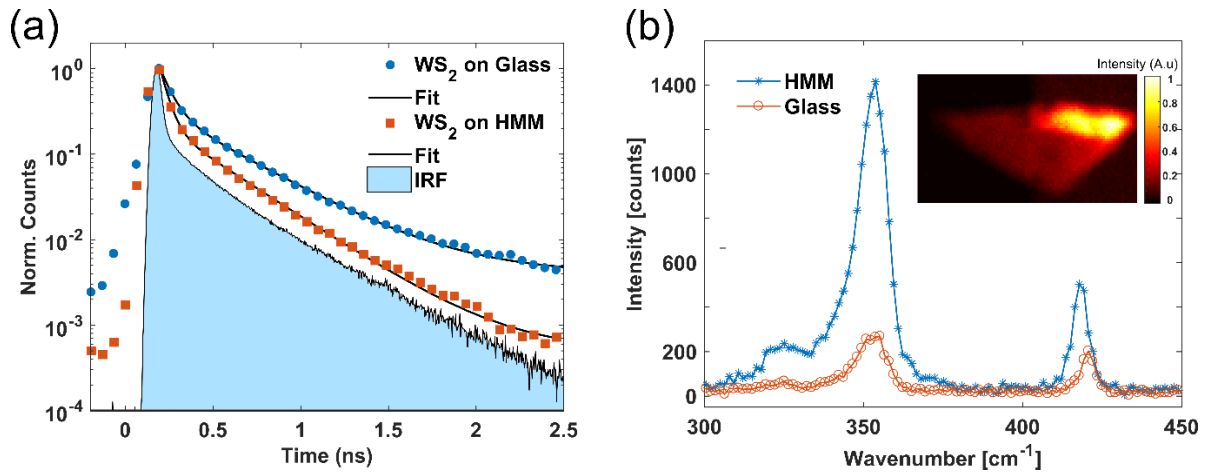

**Figure 4:** **(a)** Time-resolved PL measurements of the WS$_2$ monolayer coupled to the HMM nanoantenna (blue dots), glass substrate (black dots), and instrument response function (IRF, light blue shaded area). The corresponding fits are shown in the figure. **(b)** Surface-enhanced Raman scattering from the WS$_2$ monolayer coupled to the HMM nanoantenna. Insert in the figure shows a Raman spectral image.

The total enhancement per nanoantenna and for the array, as predicted by our FDTD simulation is ~ 300 (10×30×1) and ~30 respectively. Both are in excellent agreement with the experimentally observed enhancement.

In the calculations, Fig 3 (c), the Purcell factor per nano antenna is around ~30, corresponding to ~3 for the array, when the $WS_2$ monolayer coupled to the HMM nanoantenna array is compared to the $WS_2$ on glass. We now directly demonstrate this enhancement of radiative decay rate experimentally by measuring the exciton decay dynamics of the $WS_2$ monolayer through time-resolved photoluminescence measurements. Figure 4(a) shows the photoluminescence decay over time for the $WS_2$ monolayer on top of the HMM nanoantenna array and on top of the glass substrate. Measurements were performed using a supercontinuum laser with an excitation wavelength of 514 nm at a 80 MHz repetition rate. All the decay curves are fitted with double exponential functions, and an average of the corresponding two time constants were taken for the enhancement estimation. In the case of $WS_2$ on glass, the average decay time is 0.35 ns and the average decay rate of the $WS_2$ monolayer on top of the HMM nanoantenna is ~0.1 ns. This decrease in average total decay rate indicates the Purcell enhancement of radiative decay rate in presence of the HMM nanoantenna (see SI for full fitting parameters and additional discussion). Finally, the average decay rate of the instrument response function (IRF) is ~0.065 ns (seen as the light blue shaded area in Figure 4(a)). Neglecting the effect of the IRF, the experimentally observed radiative enhancement measured by the time resolved PL is ~3.5-fold, which is in very good agreement with the calculated value for the large area of ~3 fold.

Figure 4(b) shows the surface-enhanced Raman scattering from $WS_2$ coupled to the HMM nanoantennas as compared to the glass substrate. This is further evidence of large excitation enhancement. The Raman scattering spectrum is taken with the same configuration as that of the PL measurements and excited with a 514 nm laser. The two observed Stokes lines around 350 cm$^{-1}$ and 419 cm$^{-1}$ correspond to the $E_{2g}^1$ and $A_{1g}$ peaks of the WS$_2$. The ratio of the intensities of these two peaks indicates that the WS$_2$ is a monolayer[49]. There is a small shift observed in the case of the $A_{1g}$ peak towards a smaller wavenumber around 0.2 cm$^{-1}$. This indicates a small degree of strain induced in the monolayer around the edges of the nanoantenna. We have not seen any significant shift in the $E_{2g}^1$, which indicates there may not be any plasmonic doping in the WS$_2$ monolayer from the HMM nanoantenna. The overall enhancement is about ~5-fold over a large area and around ~50-fold per cavity (see methods). This serves as a clear indication that the excitation efficiency is enhanced due to the HMM nanoantenna cavity.

**CONCLUSIONS**

In this work, we have demonstrated the enhancement of photoluminescence (PL) emission from a 2D $WS_2$ monolayer coupled to HMM nanoantennas. The enhancement is shown via PL measurements, time resolved PL measurements and Raman measurements, and is also supported by full wave simulations. The overall PL enhancement is attributed to both excitation enhancement, and radiative decay enhancement. The enhancement in the radiative decay rate of a densely packed HMM nanoantenna array has potential applications in ultrafast nanoscale light emitters and photodetectors. Importantly, the coupling of the $WS_2$ monolayer enhances the radiative decay rate through pure spectral decoupling of radiative and non-radiative channels. This new degree of freedom in the design of HMM nanoantenna may allow us in the future to enhance the radiative rate not only from low quantum yield layers, but even from moderately high quantum yield TMDCs monolayer. We have reported up to 30-fold enhancement in PL from the overall area with a ~3.5 fold enhancement attributed to radiative decay rate enhancement. Normalizing to the area of a single nanoantenna, a ~300-fold enhancement of PL intensity is observed. We have also shown that Raman scattering intensities can be enhanced ~50-fold per single nanoantenna, and 5-fold over the whole area of the $WS_2$ monolayer which is coupled to the HMM nanoantenna array. Our approach can be extended to other TMDC monolayers with applications in ultrafast light-emitting diodes, photodetectors, and other nanophotonic devices where light-matter interactions requires overcoming the Ohmic loss of the plasmonic structure for a desired wavelength. Further enhancement in the collected PL intensity can be achieve by designing our array to have a beaming effect, such that the emission pattern matches the collection optics. Our demonstration can be considered as a new platform for light-matter-interaction applications in nonlinear optics, single-photon sources, quantum light emitters with 2D materials, and 2D materials integrated to other photonic-plasmonic structures.

**Methods:**

**Sample Fabrication:** All the samples were fabricated on top of glass substrates. First, a 200 nm thick Polymethylmethacrylate (PMMA) 950K e-beam resist from Microchem was spin-coated over the substrate. To avoid charging we have used a conductive water solvable layer (E-spacer) on top of the PMMA layer[50]. Then, the sample was patterned using a 100keV Elionix electron beam lithography with an area of 100 μm × 100 μm for each cavity array. The exposed PMMA was developed for 60 seconds using MIBK: IPA (1:3) solution. Six alternating layers consisting of $Al_2O_3$ (24 nm, three layers) and Ag (16 nm, three layers), were evaporated on top of the substrate using an e-beam evaporator at deposition rates of 0.3Å/s and 0.5Å/s respectively. Lift-off was used to obtain the designed structure by treating the sample with acetone using an ultrasonic bath for 5 minutes. Finally, a 3-4 nm $Al_2O_3$ capping layer was deposited by atomic layer deposition (ALD) to protect the silver layers from oxidization. In parallel, $WS_2$ monolayers were obtained by standard mechanical exfoliation from a $WS_2$ single crystal from HQ graphene onto a $Si/SiO_2$ (285 nm) substrate. PL measurements were used to verify the single layers. After single layer identification, the flakes were picked up from the Si substrate using a PDMS/tape/PMMA stack and transferred to the HMM nanoantenna structure by dry transfer technique using a home built 2D layer transfer setup.

**Transmission measurement:** To characterize the optical response of the sample we have performed transmission measurements by illuminating it with a white light source (tungsten-halogen lamp) through a microscope condenser lens. The transmitted light is collected by an objective lens (Nikon, 50×, NA 0.45) and directed into an Ocean Optics, Flame spectrometer through an aperture placed on the image plane for the spatial selection of the transmitted light.

**PL measurements:** Both steady-state PL and Raman scattering measurements were performed with a 514nm excitation wavelength. Both PL and Raman scattering measurements were performed using a commercial RENISHAW inVia Qontor confocal Raman microscope in reflection mode. Time-resolved photoluminescence (TRPL) measurements were performed using a home built inverted microscope in reflection mode, using a 514 nm wavelength excitation source at a repetition rate 80 MHz. The TRPL signal was coupled to a fiber that is connected to a single photon detector (SPAD, for time-resolved measurement). We used PicoHarp 300 for the photon counting electronics.

**Numerical simulations**: FDTD simulations were performed using the commercial software package FDTD Solutions, Lumerical Inc. All the simulations were performed with a mesh size of 0.5 nm. For transmission and calculation of the excitation enhancement, plane wave excitation was used. For all simulations, periodic boundary conditions along the x- and y-axis were used, with a perfectly matched layer (PML) boundary condition along the z-axis. For radiative rate enhancement simulations, a dipole source was used with an emission spectrum that matches the $WS_2$ emission spectrum. By varying the dipole position and polarization we have calculated the average enhancement of decay rates for the given system. More details about simulations are given in the supplementary materials.

**Acknowledgments:**

We acknowledge funding from the Israeli Ministry of Science and Technology and The Air Force Office of Scientific Research. CF is supported by the Carlsberg Foundation as an Internationalization Fellow. Samples were fabricated at the center for nanoscience and nanotechnology of the Hebrew university. We thank Atzmon Vakahi and Sergei Remennik for FIB and STEM measurements.

For Table of Contents Only

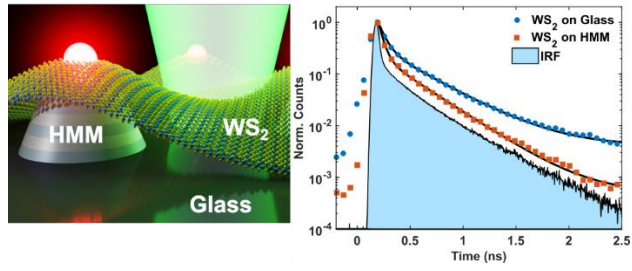